\documentclass{article}
 \usepackage[dvipdf]{epsfig}
\usepackage{color}
 
\begin{document}

\def\bw{\textbf{w}}
\def\bx{\textbf{x}}
\def\bc{\textbf{c}}
\def\bg{\textbf{g}}
\def\bk{\textbf{k}}
\def\a{{\alpha }}
\def\bt{{\beta }}
\def\be{\begin{equation}}
\def\ee#1{\label{#1}\end{equation}}
\newcommand{\ben}{\begin{eqnarray}}
\newcommand{\een}{\end{eqnarray}}
\def\d{\textsf{d} }
\def\b{\textsf{b} }


 \title{\bf Entropy, entropy flux and entropy  rate\\ of granular materials}

\author{Gilberto  M. Kremer\footnote{kremer@fisica.ufpr.br}
\\ \\
Departamento de F\'{\i}sica, Universidade Federal do Paran\'a\\
Caixa Postal 19044, 81531-990 Curitiba, Brazil}
 \date{}
 \maketitle


\begin{abstract}
{The aim of this work is to analyze the entropy, entropy flux and entropy  rate of granular materials within the frameworks of the Boltzmann equation and continuum thermodynamics. It is shown that the entropy inequality for a granular gas that follows from the Boltzmann equation differs from the one of a simple fluid due to the presence of a term which can be identified as the entropy  density rate. From the knowledge of a non-equilibrium distribution function -- valid for for processes closed to equilibrium  -- it is obtained that the entropy  density  rate is proportional to the  internal energy  density rate  divided by the temperature, while the entropy flux is equal to the heat flux vector divided by the temperature. A thermodynamic theory of a granular material is also developed whose objective is the determination of the basic fields of mass density, momentum density and internal energy density. The constitutive laws are restricted by the principle of material frame indifference and by the entropy principle. Through the exploitation of the entropy principle with Lagrange multipliers, it is shown that the results obtained from  the kinetic theory for granular gases concerning the entropy  density rate and entropy flux  are valid in general for processes close to equilibrium of granular materials, where linearized constitutive equations hold.}
\end{abstract}



\section{Introduction}

 In the last two decades a great number of papers has appeared in the
 literature concerning the theory and applications of granular
 gases. The determination of the equilibrium and non-equilibrium
 distribution functions from the Boltzmann equation was investigate
 among others by Lun {\it et al.}~\cite{Lun}, Jenkins and
 Richman~\cite{JR1,JR2}, Goldshtein and Shapiro~\cite{GS}, Brey {\it
 et al.}~\cite{BDKS} and a more complete list of the works can be found in the books~\cite{1,2,3}.
Whereas for a rarefied gas of elastic particles the collisions
conserve the mechanical energy and the gas relaxes to an equilibrium
state described by a Maxwellian distribution function, the inelastic
collisions between the particles of a granular gas transform the
translational kinetic energy into heat and there exists no equilibrium state which is
characterized by a Maxwellian distribution function. Several
remarkable properties of  granular gases are reported in the
literature, among others, the decay of the temperature of a granular
gas (the so-called Haff's law), nonhomogeneous structure formation
(cluster formation), shear instabilities (shock wave formation),
anomalous diffusion, etc.

While the  transport properties of
granular gases are almost well understood, there is a lack in the
literature, to the best of our knowledge,  concerning the discussion
of the entropy, entropy flux and entropy  rate of a granular material and the aim of this work is to analyze
these subjects within the framework of the Boltzmann equation and from the viewpoint of continuum thermodynamics.

{In Section 2 we  obtain the balance equation for the entropy density of a granular gas from the Boltzmann equation, by showing that there exists  a positive semi-definite quantity which can be identified with the  production  rate of the  entropy density. The so-called entropy inequality for a granular gas  differs from the one of a simple fluid due to the presence of a term which can be interpreted as the  entropy  density rate. Due to the energy loss of the gas particles an entropy density rate must be present in order to preserve the positiveness of the production  rate of the  entropy density. From the knowledge of the non-equilibrium distribution function -- which is  valid for processes close to equilibrium  -- the constitutive equations for the pressure tensor, heat flux vector,  internal energy  density rate (which is related to the cooling rate of a granular gas), entropy flux and  entropy density rate are calculated. It is shown  that the  entropy  density rate is proportional to the internal energy  density rate divided by the temperature, while the entropy flux is equal to the heat flux vector divided by the temperature.}

{The results obtained in Section 2 are used in Section 3 as a motivation for the  development of a thermodynamic theory for granular materials. The  objective of such theory is the determination of the basic fields of mass density, momentum density and internal energy density. The necessary field equations are based upon the balance equations of the basic fields which are closed by constitutive laws. The principle of material frame indifference and the entropy principle are used in order to restrict the generality of the constitutive equations for the internal energy density, pressure tensor, heat flux vector,  internal energy  density rate, entropy density, entropy flux and entropy  density rate. From the exploitation of the entropy principle with Lagrange multipliers we show that for processes close to equilibrium, the relationships between  the entropy  density rate and the internal energy  density rate and between the entropy flux and the heat flux vector found in Section 2 by using a restricted non-equilibrium distribution function are  valid  in general for granular materials within a linearized theory. }

\section{Kinetic theory of a rarefied granular gas}

In this section we review briefly  the main features of a granular gas which
is described by the Boltzmann equation and obtain  the balance equation for its entropy density. We consider a rarefied
granular gas free of external body forces where only binary
collisions between the particles are taken into account. Let us
denote by $m$ and $\d$ the mass and the diameter of a spherical particle and
by $(\bc, \bc_1)$ and $(\bc^\prime, \bc_1^\prime)$ the velocities of
two particles before and after collision, respectively. If
$\bg=\bc_1-\bc$ and $\bg^\prime=\bc_1^\prime-\bc^\prime$ represent
the relative velocities before and after collision, the inelastic
collisions will be characterized by the relationship
$(\bg^\prime\cdot\bk)=-\alpha (\bg\cdot\bk)$ where $0\leq\alpha\leq1$
denotes a normal restitution coefficient and $\bk$ is the
 unit collision vector which joins the centers of the two colliding spheres
 pointing from the center of the particle denoted by 1 to the
 center of the other particle without index.

The velocities before and after collision are related by the equations
 \be
 \bc^\prime=\bc+{1+\alpha\over2}(\bg\cdot\bk)\bk,\qquad
 \bc_1^\prime=\bc_1-{1+\alpha\over2}(\bg\cdot\bk)\bk.
 \ee{1}
  From the above equations it follows  the relationships which connect the relative velocities and  their modulus, namely,
 \be
 \bg^\prime=\bg-(1+\a)(\bg\cdot\bk)\bk,\qquad
 g^{\prime2}=g^2-(1-\a^2)(\bg\cdot\bk)^2.
 \ee{2a}
 Furthermore, the variation of the kinetic energy in terms of the pre- and post-collisional velocities, reads
 \be
 {m\over2}c^{\prime2}+{m\over2}c_1^{\prime2}-{m\over2}c^{2}-{m\over2}c_1^{2}
 ={m\over4}(\a^2-1)(\bg\cdot\bk)^2.
 \ee{2b}
When $\a=1$, which is the case of elastic collisions, it follows the conservation of the kinetic energy.

  A direct encounter is characterized by the pre-collisional velocities $(\bc, \bc_1)$, by the post-collisional velocities $(\bc^\prime, \bc_1^\prime)$
  and by the collision vector $\bk$. For a restitution encounter
  the pre- and post-collisional velocities are denoted by $(\bc^\ast, \bc_1^\ast)$ and $(\bc, \bc_1)$, respectively,  and the collision vector by $\bk^\ast=-\bk$.
 It is easy to verify that the relationship $\bg\cdot\bk^\ast=-\alpha(\bg^\ast\cdot\bk^\ast)=-(\bg\cdot\bk)$ holds.

In a restitution encounter the pre-collisional velocities $(\bc^\ast, \bc_1^\ast)$ are related to the post-collisional velocities $(\bc, \bc_1)$  by the equations
 \be
 \bc=\bc^\ast+{1+\a\over2}(\bg^\ast\cdot\bk^\ast)\bk^\ast,\qquad
 \bc_1=\bc_1^\ast-{1+\a\over2}(\bg^\ast\cdot\bk^\ast)\bk^\ast.
 \ee{2c}
 By using  the relationships $\bk^\ast=-\bk$ and $(\bg\cdot\bk)=-\a(\bg^\ast\cdot\bk)$ that characterize a restitution collision, the above equations can be written as
 \be
 \bc^\ast=\bc+{1+\a\over2\a}(\bg\cdot\bk)\bk,\qquad
  \bc_1^\ast=\bc_1-{1+\a\over2\a}(\bg\cdot\bk)\bk.
 \ee{2d}

For the  determination of the Boltzmann equation we have  to know the transformation of the volume elements $d{\bf c}_1^\ast\,d{\bf
c^\ast}=\vert J\vert d{\bf c}_1\,d{\bf c}$ where $\vert J\vert$ is the modulus of the Jacobian of the transformation.  By a straightforward calculation $\vert J\vert={1/\a}$ and it follows that
 \be
 (\bg^\ast\cdot\bk^\ast)\,d\bc^\ast\,d{\bf c}_1^\ast={1\over\a^2}(\bg\cdot\bk)\,d\bc\,d{\bf
 c}_1.
 \ee{2e}

 From the expression (\ref{2e}) one may infer that the Boltzmann equation for granular gases without external forces  is given by
 \be
 {\partial f \over \partial t} +c_i{\partial f \over
 \partial x_i}
 =\chi\int\left({1\over\a^2} f_1^\ast f^\ast-f_1f\right)\,\d^2\, (\bg\cdot\bk)\,d\bk\,d{\bf c}_1,
 \ee{3}
where $\chi$ is the radial equilibrium distribution,
which takes into account the correlation between the positions of
the colliding particles.

 The multiplication of the Boltzmann equation (\ref{3}) by an arbitrary
 function  $\psi(\bx,\bc,t)$ leads to
 the so-called transfer equation
 \ben\nonumber
&& {\partial \over\partial t} \int \psi f\,d{\bf c}+{\partial \over\partial x_i} \int \psi c_i f\,d{\bf c}- \int\left[{\partial \psi\over\partial t}+c_i{\partial \psi\over\partial x_i}\right]f d\bc\\\nonumber
 &=&{\chi\over2}\int\psi(\bx,\bc,t)\left({1\over\a^2} f_1^\ast f^\ast-f_1f\right) \,\d^2\, (\bg\cdot\bk)\,d\bk\,d{\bf
 c}_1\,d{\bf c}
 \\\label{4}
 &=&{\chi\over2}\int\bigl[\psi(\bx,\bc_1^\prime,t)+\psi(\bx,\bc^\prime,t)
 -\psi(\bx,\bc_1,t)-\psi(\bx,\bc,t)\bigr]
 f_1f \,\d^2\, (\bg\cdot\bk)\,d\bk\,d{\bf
 c}_1\,d{\bf c}.
 \een
 The second equality above was obtained by taking into account the relationship (\ref{2e}), by renaming the the pre-collisional velocities $(\bc^\ast, \bc_1^\ast)$  as $(\bc, \bc_1)$  and the post-collisional velocities $(\bc, \bc_1)$  as $(\bc', \bc_1')$ and by using the symmetry properties of the collision term when the two molecules are interchanged.

A macroscopic state of the granular gas is characterized by the
fields of mass density $\varrho$, hydrodynamic velocity $v_i$
and internal energy density  $\varrho\varepsilon$ defined by
 \be
 \varrho=\int mf\,d{\bf c},\qquad
  \varrho v_i=\int m c_i f\,d{\bf
 c},\qquad
  \varrho\varepsilon=\int{m\over2} C^2 f\,d{\bf c},
 \ee{7}
where $C_i-v_i$ is the so-called peculiar velocity.

 The balance  equations for the fields (\ref{7}) are obtained by choosing $\psi$ equal to $m$,
 $mc_i$ and $mC^2/2$ in the transfer equation (\ref{4}), yielding
 \ben\label{8a}
 {\partial \varrho \over\partial t} &+&{\partial \varrho v_i \over\partial x_i}=0,\\\label{8b}
  {\partial \varrho v_i\over\partial t} &+&{\partial (\varrho v_i v_j+p_{ij})\over\partial x_j}=0,\\\label{8c}
  {\partial\varrho\varepsilon\over\partial t}   &+&{\partial \left( \varrho\varepsilon v_i+q_i\right) \over\partial x_i}
 +p_{ij}{\partial v_i\over\partial v_j} +\zeta=0.
 \een
 Above, $p_{ij}$ and $q_i$ denote the pressure
 tensor and the heat flux vector, respectively, which are defined by
  \be
  p_{ij}=\int mC_iC_jfd\bc,\qquad
  q_i=\int {m\over 2}C^2C_ifd\bc.
  \ee{9}
    Furthermore, $\zeta$ is the internal energy  density rate of the granular
 gas. From the right-hand side of (\ref{4}) and the relationship (\ref{2b}) it follows that
  \be
 \zeta={\chi\d^2m(1-\alpha^2)\over8}
 \int f_1f(\bg\cdot\bk)^3d\bk\,d\bc_1\,d\bc.
 \ee{9a}
 The so-called cooling rate $\zeta^\star$ is related to the internal energy  density rate $\zeta$ by $\zeta^\star=2m\zeta/3kT\varrho$. Note that the
   internal energy  density rate vanishes for elastic collisions, i.e., for $\alpha=1$.

 By applying the Chapman-Enskog method to determine the non-equilibrium distribution
 function from the Boltzmann equation (\ref{3}) it follows~\cite{BDKS,3}
\ben\nonumber
 f&=&{\varrho\over m}\left({m\over2\pi kT}\right)^{3\over2}e^{-{m C^2\over2kT}}\Bigg\{1+\underline{{16(1-\a)(1-2\a^2)\over81-17\a+30\a^2(1-\a)}\left({15\over8}-{5\beta C^2\over2}
 +{\beta^2 C^4\over2}\right)}
 \\\label{10}
 &+&\gamma_1\left({5\over2}-{m C^2\over2kT}\right)C_i{\partial T\over\partial x_i}+
 \gamma_2C_iC_j\left({\partial v_{(i}\over\partial x_{j)}}-{1\over3}{\partial
 v_{k}\over\partial x_{k}}\delta_{ij}\right)+\gamma_3\left({5\over2}-{mC^2\over2kT}\right)C_i{\partial \varrho\over\partial
 x_i}\Bigg\}.
 \een
 Above, $k$ is Boltzmann's constant and the temperature is related to the specific internal energy through $T=2m\varepsilon/3k$. Moreover, the coefficients $\gamma_1$, $\gamma_2$ and $\gamma_3$ read
 \ben\label{11}
 \gamma_1&=&{15m\over\chi\d^2\varrho T(9+7\alpha)(\alpha+1)}\sqrt{{m\over\pi
 kT}},\qquad
 \gamma_2={-15m\over2\chi\d^2\varrho(13-\alpha)(\alpha+1)}\sqrt{{1\over\pi}\left({m\over
 kT}\right)^3},\\\label{12}
 \gamma_3&=&{300m(1-\alpha)\over\chi\d^2\varrho^2(9+7\alpha)(\alpha+1)(19-3\alpha)}\sqrt{{m\over\pi
 kT}}.
 \een
 We may observe that the non-equilibrium distribution function (\ref{12}) does  not reduces to a Maxwellian distribution function in the absence of the spatial gradients due to the presence of the underlined term. However, this term vanishes in the case of elastic collisions, i.e., for $\alpha=1$ .

 The insertion of the distribution function (\ref{10}) into the definitions of the
   pressure tensor $p_{ij}$ and heat flux
 vector $q_i$ -- given by (\ref{9}) -- and subsequent integration of the resulting equations lead to
 the linearized constitutive equations
 \be
   p_{ij}=\varrho{k\over m}T\delta_{ij}-2\mu \left({\partial
 v_{(i}\over\partial x_{j)}}-{1\over3}{\partial
 v_{k}\over\partial x_{k}}\delta_{ij}\right),\qquad
 q_i=-\kappa_0 {\partial T\over\partial x_i}-\kappa_1{\partial \varrho\over\partial
 x_i}.
 \ee{14}
 Above, the parentheses around the indexes denote the symmetric part of a tensor. Furthermore,  the coefficients of shear viscosity $\mu$, thermal conductivity
 $\kappa_0$ and $\kappa_1$ are given by
 \ben\label{15}
 \mu&=&{15\over2\chi\d^2(13-\alpha)(\alpha+1)}\sqrt{mkT\over\pi},\qquad
 \kappa_0={75\over2\chi\d^2(9+7\alpha)(\alpha+1)}{k\over m}\sqrt{mkT\over\pi},\qquad
 \\\label{16}
 \kappa_1&=&{750(1-\alpha)\over\chi\d^2(9+7\alpha)(\alpha+1)(19-3\alpha)}{kT\over\varrho}\sqrt{kT\over m\pi}.
 \een

 The linearized expression for the internal energy  density rate $\zeta$, obtained from the insertion of (\ref{10}) into (\ref{9a}) and subsequent integration of the resulting equation, yields
 \be
 \zeta_0=2{\chi k}T{\varrho^2\over m^2} \d^2 \sqrt{\frac{\pi k T}{m} } \left(1 -\alpha^2\right)\left[1+
 {3(1-\a)(1-2\a^2)\over81-17\a+30\a^2(1-\a)}\right].
 \ee{14a}
 The above expression for the internal energy  density rate is valid only  in a linearized theory where products of gradients are neglected.

 Now we may choose  $\psi(\bx, \bc, t)=-k\ln f$ into the transfer equation  (\ref{4}) and obtain
 \ben\nonumber
 {\partial\over \partial t}\int(-k\ln f)fd\bc+{\partial\over\partial
 x_i}\int (v_i+C_i)(-k\ln f)fd\bc+{\chi\d^2 k\over2}\int
 \ln{f_1^\prime f^\prime\over f_1f}f_1^\prime f^\prime(\bg\cdot\bk) d\bk d\bc_1 d\bc
 \\\label{17}
 ={\chi\d^2 k\over2}\int
 \ln{f_1^\prime f^\prime\over f_1f}\left({f_1^\prime f^\prime\over f_1f}
 -1\right)f_1f(\bg\cdot\bk) d\bk d\bc_1 d\bc.
 \een
 Equation (\ref{17}) can be written as
 \be
 {\partial \varrho s\over \partial t}+{\partial\left(\varrho sv_i+\varphi_i\right)\over\partial
 x_i}+\varsigma=\sigma,
 \ee{18}
 which can be identified with the balance equation for the entropy density $\varrho s$,  where  $\varphi_i$ is its entropy flux,  $\varsigma$ denotes the  entropy  density rate and $\sigma$ its  production rate.  The expressions for these quantities are given by
 \ben\label{19}
 \varrho s&=&-k\int f\ln f\,d\bc,\qquad
  \varsigma={\chi\d^2 k\over2}\int
 \ln{f_1^\prime f^\prime\over f_1f}f_1^\prime f^\prime(\bg\cdot\bk) d\bk d\bc_1 d\bc,
 \\\label{20}
 \varphi_i&=&-k\int C_if\ln f\,d\bc,\qquad\sigma={\chi\d^2 k\over2}\int
 \ln{f_1^\prime f^\prime\over f_1f}\left({f_1^\prime f^\prime\over f_1f}
 -1\right)f_1f(\bg\cdot\bk) d\bk d\bc_1 d\bc.
 \een
  On the basis of the relationship $(x-1)\ln x\geq0$, which is valid for all $x>0$, we may infer that the production  rate of the  entropy density is a non-negative quantity, i.e., $\sigma\geq0$.

From the insertion of the distribution function (\ref{10}) into the definition entropy flux (\ref{20})$_1$ and the integration of the resulting equation leads to the following linearized expression
\be
 \varphi_i={q_i\over T}.
\ee{21}
Hence, in a linearized theory  the entropy flux of a granular gas is equal to the heat flux vector divided by the temperature.

The determination of the  entropy  density rate (\ref{19})$_2$ and  its production  rate  (\ref{20})$_2$ is more involved. First, we may write the entropy  density rate as
\ben\nonumber
\varsigma&=&{\chi\d^2 k\over2}\int
 \ln{f_1^\prime f^\prime\over f_1f}f_1^\prime f^\prime(\bg\cdot\bk) d\bk d\bc_1 d\bc={\chi\d^2 k\over2\a^2}\int
 \ln{f_1^\prime f^\prime\over f_1f}f_1^\prime f^\prime(\bg'\cdot\bk') d\bk d\bc_1' d\bc'
 \\\label{21a}
 &=&{\chi\d^2 k\over2\a^2}\int
 \ln{f_1 f\over f_1^\ast f^\ast}f_1f(\bg\cdot\bk) d\bk d\bc_1 d\bc,
\een
by using the relationship $(\bg\cdot\bk)\,d\bc\,d{\bf c}_1={1\over\a^2}(\bg'\cdot\bk')\,d\bc'\,d{\bf
 c}'_1$ and by renaming the velocities $(\bc', \bc_1') \longmapsto(\bc, \bc_1)$   $(\bc, \bc_1)\longmapsto(\bc^\ast, \bc_1^\ast)$. In a linearized theory
\be
\ln{ff_1\over f_1^\ast f^\ast}=-{m\over2kT}(C_1^2+C^2-C_1^{\ast2}-C^{\ast2})={m\over4kT}(1-\a^2)(\bg^\ast\cdot\bk^\ast)^2={m\over4kT}\left({1-\a^2\over\a^2}\right)(\bg\cdot\bk)^2.
\ee{21b}
The insertion of (\ref{21b}) together with the distribution function (\ref{10}) and integration of the resulting equation leads to
\be
\varsigma_0={1\over\a^4}{\zeta_0\over T},
\ee{21c}
showing that the entropy density rate in a linearized theory  is proportional to the internal energy density rate.

By following the same methodology, it is easy to obtain that
\be
\ln{f'f_1'\over f_1  f }=-{m\over2kT}(C_1^{'2}+C^{'2}-C_1^{2}-C^{2})={m\over4kT}(1-\a^2)(\bg\cdot\bk)^2,
\ee{21d}
so that production  rate of the  entropy density in a linearized theory reduces to
\be
\sigma_0=\left({1\over\a^4}-1\right){\zeta_0\over T}
\ee{21e}

Note that for restitution coefficients $\a\leq1$, the internal energy density rate (\ref{14a}), the entropy density rate (\ref{21c}) and the production  rate of the  entropy density (\ref{21e})
 are always positive semi-definite quantities, i.e., $\zeta_0\geq0$, $\varsigma_0\geq0$ and $\sigma_0\geq0$, with the equal sign valid in the elastic limit, i.e., for  $\a=1$.
For restitution coefficients close to unit ($\a\longmapsto1$) the internal energy density rate (\ref{14a}) becomes
\be
\zeta_0=4{\chi k}T{\varrho^2\over m^2} \d^2 \sqrt{\frac{\pi k T}{m} } \left(1 -\alpha\right)+{\cal O}((1-\a)^2),
\ee{21f}
and the entropy density rate (\ref{21c}) and its production rate (\ref{21e}) reduce to
\be
\varsigma_0={\zeta_0\over T}+{\cal O}((1-\a)^2),\qquad \sigma_0={\cal O}((1-\a)^2).
\ee{21g}
Hence for quasi-elastic restitution coefficients the entropy density rate is equal to the internal energy density rate divided by the temperature, while the production  rate of the  entropy density
becomes a term of second order in $(1-\a)$.

\section{Thermodynamics of granular materials}

The main objective of a thermodynamic theory of granular materials is the determination of the fields of mass density $\varrho({\bf x}, t)$, velocity $v_i({\bf x},t)$ and temperature $T({\bf x},t)$ in all points of the material $\bf x$ at time $t$.
The knowledge of these fields are based on the balance equations of mass density (\ref{8a}), momentum density (\ref{8b}) and internal energy density (\ref{8c}).

However, the system of equations (\ref{8a}) through (\ref{8c}) is not closed for the determination of the basic fields $\varrho$, $v_i$ and $T$, since it is necessary to express the constitutive quantities $p_{ij}, q_i, \varepsilon$ and $\zeta$ in terms of the basic fields. The generic form of the constitutive equations which obey the principle of material frame indifference and refer  to a viscous and heat-conducting granular material is given by
\be
\left\{\varepsilon, q_i, p_{ij}, \zeta\right\}={\cal F}\left(\varrho, T, {\partial T\over\partial x_i}, {\partial \varrho\over\partial x_i},
{\partial v_{(i}\over\partial x_{j)}}\right).
\ee{22}
The solutions of the balance equations (\ref{8a}) through (\ref{8c}) for the basic fields which takes into account the  constitutive equations (\ref{22}) are called  thermodynamic processes.

The constitutive equations are restricted also by the entropy principle which states that the entropy inequality (\ref{18}) must hold for every thermodynamic process. Furthermore, the specific entropy density $s$, the entropy flux $\varphi_i$ and the  entropy  density rate $\varsigma$  are considered constitutive quantities whose dependence is the same as the ones given in  (\ref{22}), namely,
\be
\left\{s, \varphi_i, \varsigma\right\}={\cal F}\left(\varrho, T, {\partial T\over\partial x_i}, {\partial \varrho\over\partial x_i},
{\partial v_{(i}\over\partial x_{j)}}\right).
\ee{23}

In this work we are interested  in a linearized theory, so that we may write the following linearized constitutive equations
\ben\label{24}
\varepsilon&=&\varepsilon_0(\varrho,T)+\varepsilon_1(\varrho,T){\partial v_i\over\partial x_i},\qquad
s=s_0(\varrho,T)+s_1(\varrho,T){\partial v_i\over\partial x_i},\\\label{25}
\zeta &=&\zeta_0(\varrho,T)+\zeta_1(\varrho,T){\partial v_i\over\partial x_i},\qquad
\varsigma=\varsigma_0(\varrho,T)+\varsigma_1(\varrho,T){\partial v_i\over\partial x_i},\\\label{26}
q_i&=&-\kappa_0(\varrho,T){\partial T\over\partial x_i}-\kappa_1(\varrho,T){\partial \varrho\over\partial x_i},\qquad
\varphi_i=-\varphi_0(\varrho,T){\partial T\over\partial x_i}-\varphi_1(\varrho,T){\partial \varrho\over\partial x_i},\\\label{27}
p_{ij}&=&\left[p(\varrho,T)-\eta(\varrho,T){\partial v_r\over\partial x_r}\right]\delta_{ij}-2\mu(\varrho,T)\left[{\partial v_{(i}\over\partial x_{j)}}-{1\over3}{\partial v_r\over\partial x_r}\delta_{ij}\right].
\een
Note that we are dealing with a granular material where the specific internal energy density $\varepsilon$, the specific entropy density $s$, the internal energy  density rate $\zeta$ and the entropy  density rate $\varsigma$   may depend on the mass density, temperature and on the divergence of the velocity. Furthermore, in the constitutive equation for the pressure tensor, $\eta$ is the coefficient of bulk viscosity. This term does not appear in (\ref{14})$_1$ due to the fact that in Section 2 the granular material refers to a rarefied monatomic gas.

The exploitation of the entropy inequality proceeds by using the method of Lagrange multipliers~\cite{Liu,muller} and by imposing that the inequality
\ben \nonumber
{\partial \varrho s\over \partial t}&+&{\partial\left(\varrho sv_i+\varphi_i\right)\over\partial
 x_i}+\varsigma-\Lambda\left({\partial \varrho\over\partial t}+{\partial \varrho v_i\over \partial x_i}\right)
 -\Lambda_i\left({\partial \varrho v_i\over\partial t}+{\partial (\varrho v_iv_j+p_{ij})\over \partial x_j}\right)\\\label{28}
  &-&\lambda\left({\partial \varrho\varepsilon\over\partial t}+{\partial (\varrho\varepsilon v_i +q_i)\over\partial x_i}+p_{ij}{\partial v_i\over\partial x_j}+\zeta\right)\geq0,
 \een
must hold for all thermodynamic processes. Furthermore,  the Lagrange multipliers $\Lambda, \Lambda_i$ and $\lambda$ are considered functions of
\be
\left\{\Lambda, \Lambda_i, \lambda\right\}={\cal F}\left(\varrho, T, {\partial T\over\partial x_i}, {\partial \varrho\over\partial x_i},
{\partial v_{(i}\over\partial x_{j)}}\right).
\ee{29}

The insertion of the constitutive equations (\ref{24}) through (\ref{27}) into the entropy inequality (\ref{28}) leads to an inequality which is linear in the derivatives
\be
{\partial v_i\over\partial t},\qquad{\partial \varrho\over\partial t},\qquad{\partial T\over\partial t},\qquad{\partial v_{k,k}\over\partial t},\qquad{\partial v_{k,k}\over\partial x_i},\qquad{\partial T_{,j}\over\partial x_i},\qquad {\partial \varrho_{,j}\over\partial x_i},
\ee{30}
where the comma indicates the differentiation with respect to the spatial coordinates. The resulting inequality must hold for all values of the quantities (\ref{30}), so that
 the coefficients of these derivatives must vanish and we get
\ben\label{31}
 \Lambda_i=0,\qquad \varrho\left({\partial s\over\partial \varrho}-\lambda{\partial \varepsilon\over\partial \varrho}\right)-\Lambda=0,\qquad
 {\partial s\over\partial T}-\lambda{\partial \varepsilon\over\partial T}=0,\\\label{32}
  {\partial s\over\partial v_{k,k}}-\lambda{\partial \varepsilon\over\partial v_{k,k}}=0,\qquad
  {\partial \varphi_{(i}\over\partial T_{,j)}}-\lambda{\partial q_{(i}\over\partial T_{,j)}}=0,\qquad
  {\partial \varphi_{(i}\over\partial \varrho_{,j)}}-\lambda{\partial q_{(i}\over\partial \varrho_{,j)}}=0.
 \een
Furthermore, there remains the following residual inequality
 \ben\label{33}
 \left[{\partial \varphi_i\over\partial \varrho}-\lambda{\partial q_i\over\partial \varrho}\right]{\partial \varrho\over\partial x_i}+
 \left[{\partial \varphi_i\over\partial T}-\lambda{\partial q_i\over\partial T}\right]{\partial T\over\partial x_i}+\varsigma-\lambda \zeta-
 \Lambda\varrho{\partial v_i\over\partial x_i}-\lambda p_{ij}{\partial v_i\over\partial x_j}\geq0.
 \een

Now we proceed to analyze the conditions (\ref{31}) and (\ref{32}). First the insertion of the constitutive equations (\ref{26}) for the heat flux vector and entropy flux into the equations (\ref{32})$_{2,3}$ leads to
\be
\varphi_i=\lambda q_i,\qquad\hbox{where}\qquad \lambda=\lambda(\varrho,T).
\ee{34}
Next from equations (\ref{31})$_3$, (\ref{32})$_1$ and (\ref{34})$_2$ it follows that
\be
{\partial(s-\lambda\varepsilon)\over \partial v_{k,k}}=0,\qquad {\partial(s-\lambda\varepsilon)\over \partial T}=-\varepsilon{\partial \lambda\over\partial T}.
\ee{35}
From the above equations we may conclude that the specific internal energy density $\varepsilon$ and the specific entropy density $s$ do not depend on the divergence of the velocity and in a linearized theory the constitutive equations (\ref{24}) reduce to
\be
\varepsilon=\varepsilon_0(\varrho,T),\qquad
s=s_0(\varrho,T).
\ee{36}
These results are the same as those found for a simple fluid~\cite{muller,Liu2}.

The next step is to build the differential of the specific entropy density $s$, which by the use of the relationships (\ref{31})$_{2,3}$, yields
\be
ds={\partial s\over\partial T}dT+{\partial s\over\partial \varrho}d\varrho=\lambda\left(d\varepsilon+{\Lambda\over\varrho\lambda}d\varrho\right).
\ee{37}
The comparison of the above equation with the Gibbs equation of thermodynamics
\be
ds={1\over T}\left(d\varepsilon-{p\over\varrho^2}d\varrho\right),
\ee{37a}
leads to the identification of the Lagrange multipliers
\be
\lambda={1\over T},\qquad \Lambda=-{p\over T\varrho}.
\ee{38}

From (\ref{34})$_1$ we get that in a linearized theory  the entropy flux is equal to the heat flux vector divided by the temperature, i.e., $\varphi_i=q_i/T$, which is the same result as the one of the kinetic theory of rarefied granular gases.

For the exploitation of the residual inequality  we insert the constitutive equations (\ref{24}) through (\ref{27}) and the relationships (\ref{38}) into the residual inequality (\ref{33}) and obtain
\be
{1\over T^2}\left(\kappa_0{\partial T\over\partial x_i}+\kappa_1{\partial \varrho\over\partial x_i}\right){\partial T\over\partial x_i}+\varsigma_0+\varsigma_1{\partial v_i\over\partial x_i}-{1\over T } \left(\zeta_0+\zeta_1{\partial v_i\over\partial x_i}\right)+{\eta\over T}{\partial v_i\over\partial x_i}{\partial v_j\over\partial x_j}+2{\mu\over T}{\partial v_{\langle i}\over\partial x_{j\rangle}}{\partial v_{\langle i}\over\partial x_{j\rangle}}\geq0.
\ee{39}
Above,  we have introduced the  velocity gradient deviator defined by
\be
{\partial v_{\langle i}\over\partial x_{j\rangle}}={\partial v_{(i}\over\partial x_{j)}}-{1\over3}{\partial v_r\over\partial x_r}\delta_{ij}.
\ee{40}

As it was pointed out in the last section, the production rate of the entropy density (\ref{21e}) does not vanish  unless  $\a=1$, which correspond to the elastic limit. In the elastic limit  the internal energy density rate and the entropy density rate vanish and the granular material reduces to a single fluid. Hence, in a state where the spatial gradients vanish  we get from the residual inequality (\ref{39}) that $\varsigma_0\geq\zeta_0/T$, with the equality sign valid only when the production  rate of the  entropy density vanishes, i.e., in the elastic limit. As was expected, this conclusion is the same as the one found from a kinetic theory of rarefied granular gases (see (\ref{21c}), (\ref{21e}) and their approximations (\ref{21g}) when $\a\longmapsto1$). Furthermore, the residual inequality (\ref{39}) is linear in the divergence of the velocity and the coefficient of this term must vanish in order to preserve the inequality and it follows that $\varsigma_1=\zeta_1/T$. Hence, as in the kinetic theory of rarefied granular gases, if the production  rate of the  entropy density could be consider as a term of small order, the entropy  density rate in a linearized theory is equal to the internal energy  density rate  divided by the temperature, i.e., $\varsigma=\zeta/T$ when $\sigma\approx0$. However, note that in the kinetic theory of gases this conclusion is valid only when the restitution coefficient is close to one, which corresponds to a quasi-elastic collision. It is worth to call attention to the fact that the entropy  density rate and the  internal energy  density rate for granular materials in a linearized theory depend also on the divergence of the velocity.

It is not possible to extract more information on the coefficients of shear and bulk viscosities and thermal conductivity from the residual inequality (\ref{39}), since we have two scalar constitutive quantities -- namely the internal energy density rate $\zeta$ and  the entropy density rate $\varsigma$ -- and we have considered linear representations for both. By taking into account quadratic terms for these two quantities, the exploitation of the residual inequality leads to results that are not remarkable to discuss here.

It is interesting to analyze the case of a simple fluid where the   entropy  density rate and the  internal energy  density rate are absent. In this case we conclude from the residual inequality that the coefficients of shear and bulk viscosities are non-negative, i.e., $\mu\geq0$ and $\eta\geq0$, and that
$\pmatrix{\kappa_0&{\kappa_1/2}\cr
{\kappa_1/2}&0}$
is a positive semi-definite matrix. This last result implies that the coefficient $\kappa_2$ must vanish for a simple fluid and   Fourier's law reduce to ${\bf q}=-\kappa_0{\nabla} T$. This conclusion is the same as that obtained by Liu~\cite{Liu2} by analyzing a simple fluid.

\section{Final Remarks}

{It is important to call attention to the fact that the relationships obtained from the thermodynamic theory in Section 3 are not restricted to rarefied monatomic granular gases as in Section 2 but also valid for granular materials in general, i.e., monatomic, polyatomic and real gases. Indeed, in the thermodynamic theory there is no dependence of the constitutive equations on the restitution coefficient, the pressure tensor has a non-vanishing bulk viscosity term and the entropy  density rate as well as the internal energy  density rate  depend on the divergence of the velocity. The  dependence of pressure tensor, the entropy  density rate and the internal energy  density rate on the divergence of the velocity could be achieved within the framework of the kinetic theory of polyatomic or dense granular gases, since their non-equilibrium distribution functions depend on the divergence of the velocity.  }

\section*{Acknowledgments} I would like to thank Professor Andr\'es Santos for valuables suggestions and the Conselho Nacional de Desenvolvimento Cient\'{i}fico e Tecnol\'ogico (CNPq) for  financial support.


\begin{thebibliography}{99}

 \bibitem{Lun} C. K. K. Lun, S. B. Savage, D. J. Jeffrey and N.
 Chepurniy, ``Kinetic theories for granular flow: inelastic particles in Couette-flow
 and slightly inelastic particles in a general flowfield'', J. Fluid Mech. {\bf 140}, 223 (1984)

 \bibitem{JR1}   J. T. Jenkins and M. W. Richman,
``Grad's 13-Moment system for a dense gas of inelastic spheres''
Arch. Ration. Mech. Anal. {\bf87}, 355 (1985)


 \bibitem{JR2} J. T. Jenkins and M. W. Richman,
``Kinetic-theory for plane flows of a dense gas of identical, rough,
inelastic, circular disks'' Phys. Fluids {\bf28}, 3485 (1985)


 \bibitem{GS}  A. Goldshtein and M. Shapiro, ``Mechanics of collisional motion of granular-materials.
 1. General hydrodynamic equations'', J. Fluid Mech. {\bf 282}, 75 (1995)


 \bibitem{BDKS} J. J. Brey, J. W. Dufty, C. S. Kim and A. Santos,
     ``Hydrodynamics for granular flow at low density'',
     {Phys. Rev. E} {\bf 58}, 4638 (1998).

 \bibitem{1} T. P\"oschel and S. Luding (Editors) \emph{ Granular Gases}
 (Springer-Verlag, Berlin, 2001).

\bibitem{2} T. P\"oschel and N. V. Brilliantov (Editors) \emph{ Granular Gas Dynamics}
 (Springer-Verlag, Berlin, 2003).


\bibitem{3} N. V. Brilliantov and T. P\"oschel \emph{ Kinetic Theory of Granular Gases}
  (Oxford University Press, Oxford, 2004).

\bibitem{Liu} I-S. Liu, ``Method of Lagrange multipliers for the exploitation of the entropy principle", Arch. rational Mech. Anal. {\bf46}, 131 (1972).

\bibitem{muller} I. M\"uller, \emph{Thermodynamics} (Pitman, London, 1985).

\bibitem{Liu2} I-S. Liu, ``On Fourier's law of heat conduction", Continuum Mech. Thermodyn. {\bf2}, 301 (1990).

 \end{thebibliography}
\end{document}